\documentclass [12pt]{article}
\usepackage {graphicx}

\begin{document}
\begin{center}
Stock mechanics: predicting recession in S{\&}P500, DJIA, and NASDAQ
\end{center}

\begin{center}
\c{C}aðlar Tuncay
\end{center}

\begin{center}
Department of Physics, Middle East Technical University
\end{center}

\begin{center}
06531 Ankara, Turkey
\end{center}

\begin{center}
caglart@metu.edu.tr
\end{center}

\textbf{Abstract}

An original method, assuming potential and kinetic energy for prices and 
conservation of their sum is developed for forecasting exchanges. 
Connections with power law are shown. Semiempirical applications on 
S{\&}P500, DJIA, and NASDAQ predict a coming recession in them. An emerging 
market, Istanbul Stock Exchange index ISE-100 is found involving a potential 
to continue to rise.



\textbf{1. Introduction}

As is well known, prices, market indices evolve from one mood to another in 
time. They may be in calm oscillatory fluctuation mood for some time, then 
they may increase (or decrease), then again turn into a new oscillatory 
period, then a crash or a crisis may come, and finally, the current era 
closes up. Then a new one takes stage; new conditions and new formations 
take place, so forth ad infinitum. The time terms between long range drastic 
changes are named as era in the present work, and comparatively medium or 
short range characteristic time periods in eras are cited as epoch. The 
whole past and future history of exchanges may be far from obeying 
analytical functions. Time series may be predicted epoch by epoch, as 
presently proposed, in terms of some simple analytical functions. These 
equations of motion for prices are derived in the next section and power law 
connections are displayed in the third one. In the fourth section three 
indices S{\&}P500, DJIA, and NASDAQ from New York Stock Exchange (NYSE) and 
ISE-100 from Istanbul Stock Exchange (ISE) are predicted by the present 
method. Last section is devoted to conclusion.

\newpage 
\textbf{2. Potential and kinetic energies and conservation of their sum}

A potential energy (taking mass as unity, see $^{NOTE }$\footnote{$^{NOTE }$ 
Some kind of inertia as ``resistance'' against any impact to move the price 
do exist in time charts of shares.[1-4, and references therein]. Yet, not 
any reasonable ``force equation'' connecting inertia and accelaration of 
stock prices is proposed. In any case, whether massless potential and 
kinetic energies are defined, or they are defined per mass, or even mass is 
taken as unity in them will not be worth much in the present scheme.\par }) 
in terms of difference in prices ($\chi =\chi $(t)) may be defined as

U($\chi $ -- $\chi _{av})$ = h ($\chi $ -- $\chi _{av})^{\alpha }$ , 
(1)

\noindent
where h and $\alpha $ are some price independent parameters designating the 
current epoch, and $\chi _{av}$ is some time average of prices, which 
defines the zero-potential-level. Eq. (1) describes a kind of 
``gravitational'' potential (as on the earth surface) for $\alpha $=1, where 
h becomes the gravitational (anti-gravitational) constant (g) with 0<h=g 
(0>h=--g). Whereas for $\alpha $=2 we have a spring-mass potential energy 
with a spring constant (Hooke constant) equal to 2h. It is worth to 
underline that, the parameters of Eq. (1) i.e., h, $\alpha $, and $\chi 
_{av}$ may differ from one epoch to the other for any share and also from 
one share to another in any epoch. Note that, potential energy in Eq. (1) 
satisfies a power law for any $\alpha $, as discussed to some extend in the 
next section.

Moreover, again by taking mass as unity, one may define kinetic energy for 
prices as

K = $\raise.5ex\hbox{$\scriptstyle 1$}\kern-.1em/ 
\kern-.15em\lower.25ex\hbox{$\scriptstyle 2$} $v$^{2}$ , (2)

\noindent
where v is the usual speed i.e., v(t)=d$\chi $(t)/dt. The dimensional unit 
of K may be taken as (local currency unit/time)$^{2}$ for shares, e.g., 
($\not{c} $/day)$^{2}$ or ({\$}/day)$^{2 }$in USA. For indices lcu may be 
kept in the units or it may be substituted by ``value''. Potential energy of 
Eq. (1) will obviously have the same unit as (lcu/time)$^{2}$, and for 
$\alpha $=1 the factor h will have the unit (lcu/time$^{2})$, where lcu 
stands for local currency unit. For $\alpha $=2, h will have the unit of 
(time$^{\mbox{--}2})$ i.e., frequency squared.

We may assume conservation of the sum of potential and kinetic energies, as 
long as friction forces, damping etc. are negligible,

U + K = h($\chi $ -- $\chi _{av})^{\alpha }$ + 
$\raise.5ex\hbox{$\scriptstyle 1$}\kern-.1em/ 
\kern-.15em\lower.25ex\hbox{$\scriptstyle 2$} $v$^{2}$ = E = constant . (3)

Differentiation of Eq. (3) with respect to time, for $\alpha $=1 yields

\noindent
h(d$\chi $/dt) + v(dv/dt) = 0 , (4)

\noindent
from which, after substituting v=d$\chi $/dt and a=dv/dt, one may obtain the 
familiar equation of motion for azimuthal rises and falls as in classical 
mechanics

$\chi $(t$_{m})=\chi _{0}$ + v$_{0}$ t$_{m}$ + 
$\raise.5ex\hbox{$\scriptstyle 1$}\kern-.1em/ 
\kern-.15em\lower.25ex\hbox{$\scriptstyle 2$} $ h t$_{m}^{2}$ , (5 )

\noindent
where $\chi _{0}$, and v$_{0}$ designate initial price and speed, 
respectively. Time t$_{m}$ runs over exchange process days and may be set to 
zero at the beginning of any epoch. For $\alpha $=2, Eq. (1) yields 
oscillations. In the expansion of $\chi _{av}$(t$_{m})=\chi 
_{av}$(t$_{m}$=0)+v$_{av}$t$_{m}$ sign and magnitude of v$_{av}$ indicates 
the up, down or horizontal character of oscillatory trends;

$\chi $(t$_{m})=\chi _{av}$(t$_{m}$=0) + v$_{av}$t$_{m}$ + Asin(wt$_{m}$ 
+ $\Phi )$ , (6)

\noindent
where A, w, and $\Phi $ is the usual amplitude, angular frequency (here, 
(2h)$^{\raise.5ex\hbox{$\scriptstyle 1$}\kern-.1em/ 
\kern-.15em\lower.25ex\hbox{$\scriptstyle 2$} })$, and phase, respectively. 
They are observed in general to have some medium range time periods about 
some ten days or so, and fade away after a few (two, or three) full periods.

\textbf{3. Potential energy, power law, and log-periodic equations of 
motion}

The form of potential energy in Eq. (1) satisfies a power law, which is 
known in physics for a long time as effective subject, utilized especially 
to express critical phenomena in statistical mechanics. Some special forms 
of power law appears outside the physical fields as Pareto's Law and Zipf's 
Law.[6, 7]. It is also utilized in seismic predictions for the rupture 
times.[8, 9]

Power law states that if the argument x of any observable O(x) is scaled by 
some $\lambda $, (i.e., if x$\prime $/x=$\lambda )$ and if O(x$\prime 
)$/O(x)=$\zeta $, then O(x)=x$^{\alpha }$ is a solution with $\lambda 
^{\alpha }=\zeta $ and $\alpha $=log$\zeta $/log$\lambda $. Note that, 
$\lambda $ and $\zeta $ are independent of x and the relative value of the 
observable at two different scales depend only on the ratio of the scaling 
parameters. This is the way scale invariance is associated to 
self-similarity and criticality. Note also that there is no condition on 
$\alpha $ to be real. Incorporating $\lambda ^{\alpha }$/$\zeta 
$=1=exp(i2$\pi $m)[10, 11], where m is any integer, one may generalize the 
standard scaling O(x)=x$^{\alpha }$ to a log-periodic one, O(x)=x$^{\alpha 
}$P(logx/log$\lambda )$, where P is a function of period 1. Fourier 
expansion of P can be performed to obtain the most general form of the 
relevant function. (For detailed and complete treatment, and for various 
applications see [8-31].) For the sake of simplicity, one may take into 
account only the first Fourier term;

O($\tau ) \quad  \approx $ (1--$\tau )^{\alpha }${\{}d$_{0}$ + 
d$_{1}$cos[2$\pi \Omega $ln(1--$\tau )+\Psi $]{\}} , (7)

\noindent
where, $\tau $ stands for t/T$_{c}$ and t is the general independent time 
variable, and T$_{c}$ is the crtitical time; $\alpha $=ln$\zeta $/ln$\lambda 
$, $\Omega $= 1/ln$\lambda $, and $\Psi _{n}$ is some general phase term. 
What is in Eq. (7) relevant to time series of shares and indices is that, 
near the crash (which is considered as a failure time for the 
log-periodicity) the frequency of the oscillations and the volatility 
increases, which can be considered as one of the hall-marks of the coming 
crash.

Let's approximate the (1--$\tau )^{\alpha }$ factor and ln(1-$\tau )$ by 
(1--$\alpha \tau )$ and (-$\tau )$ for $\tau  \cong $0, i.e. much before 
(and by t $ \to $ --t, much after) the critical time. After some simple 
mathematical manipulations Eq. (7) can be written as

O($\tau ) \quad  \approx $ D$_{0}$ + D$_{1}\tau $ + D$_{2}$cos(2$\pi \Omega 
\tau +\Psi \prime )$ , (8)

\noindent
where, the constants D$_{0}$, D$_{1}$, D$_{2}$ can be calculated out of 
d$_{0}$, d$_{1, }$and the others of Eq. (7). Close similarity between Eqs. 
(8) and (6) is a consequence of Eq. (1) obeying power law.

\textbf{4. Applications}

The three NYSE indices S{\&}P500, DJIA, and NASDAQ[33] are extensively 
studied in literature especially within the formalism of power law. For 
similar log-periodic predictions performed on ISE-100[34] see [27]. In all 
of these indices (as in many other world markets) a severe crash dated about 
the year of 2000 is common. The present state of the same indices will be 
investigated utilizing the original method of stock mechanics.

\textit{The index of S{\&}P500}

A crucial feature in S{\&}P500 (Fig. 1) is the almost symmetric behavior of 
time series about the critical point near 01.Sept.2000 with the close value 
of 1520.77. Secondly, starting with the beginning of 1995, oscillatory 
periods decrease as closes climb to the climax. Afterwards closes start to 
recede and high frequency oscillations turn back into low frequency ones 
with increasing amplitudes. Thirdly, with the linear price axis, the time 
series may be fitted as a first order approximation by partial straight 
lines. Then simple analytical functions may be utilized for each epoch (j=1, 
2, 3, 4) as $\chi _{j}$(t$_{m})=\chi _{0j}$ + v$_{j}$t$_{m}$, 
corresponding to a constant potential energy, i.e. $\alpha $=0, and h 
arbitrary in Eq. (1). In this picture the 1987, 1990, 1998 crashes do seem 
as normal fluctuations, as well as the others after 2000.

For $\alpha $=1 in Eq. (1), the second order expression of Eq. (5) delivers 
very interesting results for the two epochs; one from the beginning of 1997 
to the end of 2002, and the second after 2002 till the present time 
(May.2005), see Fig. 1. By a simple least square fit (lsf) to daily 
data[33], the gravity comes about the same for both of the pronounced epochs 
as h = -- 0.001101 lcu/day$^{2}$. The initial (shooting) speed is found to 
be 1.73 lcu/day for the first epoch and 1.11 lcu/day for the next one. 
Imagining the close values as height of a particle shot up in the given 
gravitational field; the particle first rises till the climax, then falls 
down and hits the ground at an elevation of 663 lcu. Afterwards it bounces 
back with a smaller speed and rises till a lesser height of 1219 lcu. 
Therefore, during the collision it looses its total energy by 59{\%} and the 
collision is inelastic. The maximum height in any epoch can be calculated 
utilizing the relation $\chi $ -- $\chi _{0}$ = v$_{0}^{2}$/(2h). It is 
worth to forecast that March.2005 values are local maximum for S{\&}P500, 
and a recessional correction may be expected till the level of 800 back, 
within the coming 500 days. The pronounced parameters are listed in Table 1 
for S{\&}P500 as well as for the other indices.

\textit{The index of DJIA}

DJIA has similar features as pronounced above for S{\&}P500, see Fig. 2. 
Focusing on quadratic behavior (Eq. 5), gravity again comes out as common 
for both of the token epochs, before and after the beginning of 2003 (Fig. 
2.), where h = -- 0.00606 lcu/day$^{2}$. Hitting speed is 10.17 lcu/day, and 
bouncing back speed is 8.43 lcu/day, corresponding to 31{\%} loss in total 
energy. So the March.2005 height is considerably close the historical top of 
Jan.2000. Again, in about 500 days, DJIA is forecasted to recede back to 
8000's.

\textit{The index of NASDAQ}

Nasdaq has the most complicated appearance of all the NYSE indices studied 
here. Yet, it displays very many similarities with S{\&}P500 and DJIA, Fig. 
3. Moreover a common gravity comes out for the two epochs following the 
beginning of 1997 and separated by the beginning of 2003 (Fig. 3.), where h= 
-- 0.0041 lcu/day$^{2}$. The hitting and bouncing back speeds are 6.08 
lcu/day and 3.10 lcu/day, respectively. Then, the loss in total energy is 
75{\%}. So, as expected, the March.2005 heights are quite below the 
historical maximum. Consecutively, NASDAQ is also forecasted to recede back 
to 1400's at least, within the next one and a half year.

\textit{The index of ISE-100}

ISE is a well known world emerging market, and comparing to the NYSE 
indices, ISE-100 has many more different aspects than similar ones. A 
log-linear era (lasting about 20 years from the beginning on) has closed by 
the 2000 crash. Afterwards a recession with 30{\%} loss in a year has taken 
place. Between Jan.2000 and Jan.2004, recession epoch has been completed and 
transition to a new up trend has already taken place.

Within the pronounced epoch, ISE-100 displays a dishlike form, and as can be 
seen in Fig. 4. a. there exist anti-gravity with h= 0.001811 lcu/day$^{2}$. 
The work done by this constant anti-gravity results in increasing the total 
energy, day by day. So, one may expect ISE-100 to continue to rise with some 
possible decorative up and down fluctuations. It is hard to forecast the 
time of departure from Eq. (5) and solid curve in Fig. 4. a.; yet, it seems 
that it lies in the far future.

On the other hand, it can be observed that, at the bottom of the Jan.2000 
and Jan.2004 epoch, the trend is horizontal. Meanwhile, many oscillations of 
type Eqs. (6) and (8) may be expected to exist in ISE-100 and in many ISE 
shares. In Figs. 4. b. and 4. c., two typical oscillatory epochs with 
different time domains are exemplified, where time axis is weekly and daily, 
respectively. The corresponding mechanical parameters are listed in Table 2. 
For better fits one may take into account many coupled smaller 
spring-masses.

\textbf{5. Conclusion}

The present analytical method can be applied to shares as well. In general 
there exists a wide diversity of epochs in world markets, in which the 
present analytical functions can safely be applied. For more elaborate epoch 
formations, some more complicated functional forms may be tried in Eq. (1). 
Or, the solutions of the present form for non-integer fractal powers of 
$\alpha $ may be taken into account. Yet, mismatches between the real and 
calculated values may always exist, due to unpredictability character of 
short-range fluctuations about longer-range ones. It is obvious that, such 
analytical approaches may be used together with the traditional approaches 
for better prediction of the markets.

\textbf{Acknowledgement}

The author is thankful to Dietrich Stauffer for his encouregement to write 
the present paper, friendly discussions and corrections, and supplying some 
references.

\textbf{References}

[1] P. Gopikrishnan \textit{et al.}, Phys. Rev. E \textbf{62}, 4 (2000).

[2] J-P. Bouchaud, R. Cont, \textit{Preprint}: cond-mat/9801279.

[3] P. Gopikrishnan \textit{et al.}, Physical Review E \textbf{60}, 5305 (1999).

[4] R. Cont, J-P Bouchaud, Macroeconomic Dynamics \textbf{4}, 170 (2000).

[5] D. C. Giancoli, ``\textit{Physics for Scientists {\&} Engineers}'', 3$^{rd}$ ed. Prentice Hall. pp. 375.

[6] V. Pareto ``\textit{Cours d''economie politique reprinted as a volume of Oeuvres Compl`etes}''

(Geneva, Droz, 1965).

[7] G. Zipf, ``\textit{Human Behavior and the Principle of Last Effort}'' (Cambridge, MA: Addison-

Wesley, 1949).

[8] H. Saleur\textit{ et al.}, J.Geophys.Res. \textbf{101}, 17661 (1996).

[9] D.J. Varnes and C.G. Bufe, Geophys. J. Int. \textbf{124}, 149 (1996).

[10] J. A. Feigenbaum and P. G.O. Freund, arXiv:cond-mat/9509033.

[11] D. Sornette\textit{ et al.}, J. Phys. I Fr. \textbf{6}, 167 (1996). \textit{Preprint}: 
arXiv:cond-mat/9510036.

[12] D. Sornette, Phys. Rep. \textbf{297}, 239 (1998).

[13] J-P. Bouchaud, Quant. Fin. \textbf{1}, 105 (2001).

[14] X. Gabaix\textit{ et al.}, Nature \textbf{423}, 267 (2003).

[15] X. Gabaix Quarterly Journal of Economics 114 (3), 739 (1999).

[16] Y. Huang \textit{et al.}, Europhysics Letters \textbf{41}, 43 (1998). \textit{Preprint}: 
arXiv:cond-mat/9612065.

[17] W. I. Newman\textit{ et al.}, Phys. Rev. E \textbf{52}, 4827 (1995).

[18] S. Dro¿d¿\textit{ et al.}, Eur. Phys. J. B \textbf{10}, 589 (1999).

[19] S. Dro¿d¿\textit{ et al.}, Physica A \textbf{324}, 174 (2003).

[20] J-P. Bouchaud, J. Kockelkoren, M. Potters. \textit{Preprint} 
http://xxx.lanl.gov/abs/cond-

\noindent
mat/0406224.

[21] C. Tannous, and A. Fessant, \textit{Preprint }arXiv:physics/0101042.

[22] D. Sornette, and A. Johansen, \textit{Preprint}: arXiv:cond-mat/9704127.

[23] A. Johansen, O. Ledoit, D. Sornette, Int. J. of Theoretical and Applied 
Finance 3, 219

(2000).

[24] W-X. Zhou, and D. Sornette, Physica A \textbf{330}, 584, \textit{Preprint} 
arXiv:physics/0301023.

[25] D. Sornette and W-X. Zhou, Quant. Fin \textbf{2}, 468 (2002).

[26] A. Johansen, O. Ledoit, D. Sornette, Int. J. of The. and Appl. Finance 
\textbf{3}, 219 (2000).

[27] W-X. Zhou, D. Sornette, Physica A \textbf{330}, 543, \textit{Preprint }:arXiv:cond-mat/0212010

[28] J. Laherr\`{e}ere and D. Sornette, Eur. Phys. J. B \textbf{2}, 525 
(1998).

[29] A. Johansen, and D. Sornette, Eur. Phys. J. B \textbf{18}, 163 (2000).

[30] For many other articles of D. Sornette see also several issues of the 
journal Eur. Phys. J.

B. and search \textit{Preprint}: http://xxx.lanl.gov/abs/cond-mat/

[31] J-P. Bouchaud, Quant. Fin. \textbf{1}, 105 (2001).

[32] For detailed information about NYSE shares and indices, URL: 
http://biz.yahoo.com/i/.

[33] For detailed information about ISE.

URL: http.//.www.imkb.gov.tr/sirket/sirketler{\_}y{\_}2003.thm.

Table 1 Physical parameters of S{\&}P500, DJIA, NASDAQ, and ISE-100 for 
$\alpha $=1 in Eq. (1). For the time domains, see text. (lcu) in some units 
stand for local currency unit.

S{\&}P500 DJIA NASDAQ ISE-100

\noindent
h (day$^{ - 2})$ -0.001101 -0.00606 -0.004119 +0.001811

\noindent
v$_{01}$ (lcu/day) 1.73 10.17 6.08 -7.521

\noindent
v$_{02}$ (lcu/day) 1.11 8.43 3.10 not present

\noindent
energy loss ({\%}) 59 31 75 not present

Table 2 Physical parameters of ISE-100 for $\alpha $=2 in Eq. (1). (lcu) in 
some units stand for local currency unit.

2001-2003 2004

\noindent
h (day$^{ - 2})$ 0.051911 0,093884

\noindent
v$_{av}$ (lcu/day) 14 18,23

A (lcu) 2950 880

\textbf{Figure captions}

Fig. 1. The Sept.2000 crash in S{\&}P500 is described as a second order 
approximation by an azimuthal rise and fall in a gravity h= --0.001101 
(lcu/day$^{2})$, where the initial (shooting) speed is v$_{01}$=1.73 
(lcu/day). The price fall down after the maximum height and inelastically 
bounces back with v$_{02}$= 1.11 (lcu/day) in the same gravity, and rises up 
to 1200's in accordance with the expression $\chi $ -- $\chi _{0}$ = 
v$_{0}^{2}$/(2h). A recession, back to 800's is predicted within the 
coming 500 days.

Fig. 2. The excursion of DJIA about the Apr-Sep.2000 climax is described as 
a second order approximation by an azimuthal rise and fall of the price in a 
gravity h= --0.00606 (lcu/day$^{2})$. The initial (shooting) speed at the 
beginning of 1995 is v$_{01}$= 10.17 (lcu/day). The price fall down after 
the maximum height and inelasticly bounces back with v$_{02}$= 8.43 
(lcu/day) in the same gravity, and rises up to 11000's in accordance with 
the expression $\chi $ -- $\chi _{0}$ = v$_{0}^{2}$/(2h). A recession, 
back to 8000's and below is predicted within the coming 500 days.

Fig. 3. NASDAQ's azimuthal motion beginning with the year of 1995 is 
described by a gravity h = --0.004119 (lcu/day$^{2})$ and initial (shooting) 
speed of v$_{01}$=6.08 (lcu/day). The inelastic bouncing speed in the same 
gravity is v$_{02}$=3.10 (lcu/day). A recession, from the present heights 
back to 1200's is predicted within the coming one and a half year.

Fig. 4. a. The epoch begun with the beginning of 2000 has an anti-gravity 
h=0.001811 (lcu/day$^{2})$. Rise is predicted to last till the departure of 
the price from the the solid curve.

Fig. 4. b. Long term oscillatory motions of ISE-100 within the dishlike 
epoch, corresponding to $\alpha $=2 in Eq. (1). The horizontal axis is 
weekly in time and h=0.051911 (week$^{ - 2})$. Oscillation fades away after 
three full periods (here, about two years and a half), as usual.

Fig. 4. d. Short term oscillatory motions of ISE-100 within the year of 2004 
with h= 0,093884 (day$^{ - 2})$. Oscillation fades away after three full 
periods (here, about two monts or so), as usual. (Notice the relative 
increase in volume at dips of oscillations.)

\textbf{Figures}

\begin{figure}[htbp]
\centerline{\includegraphics[width=5.87in,height=5.17in]{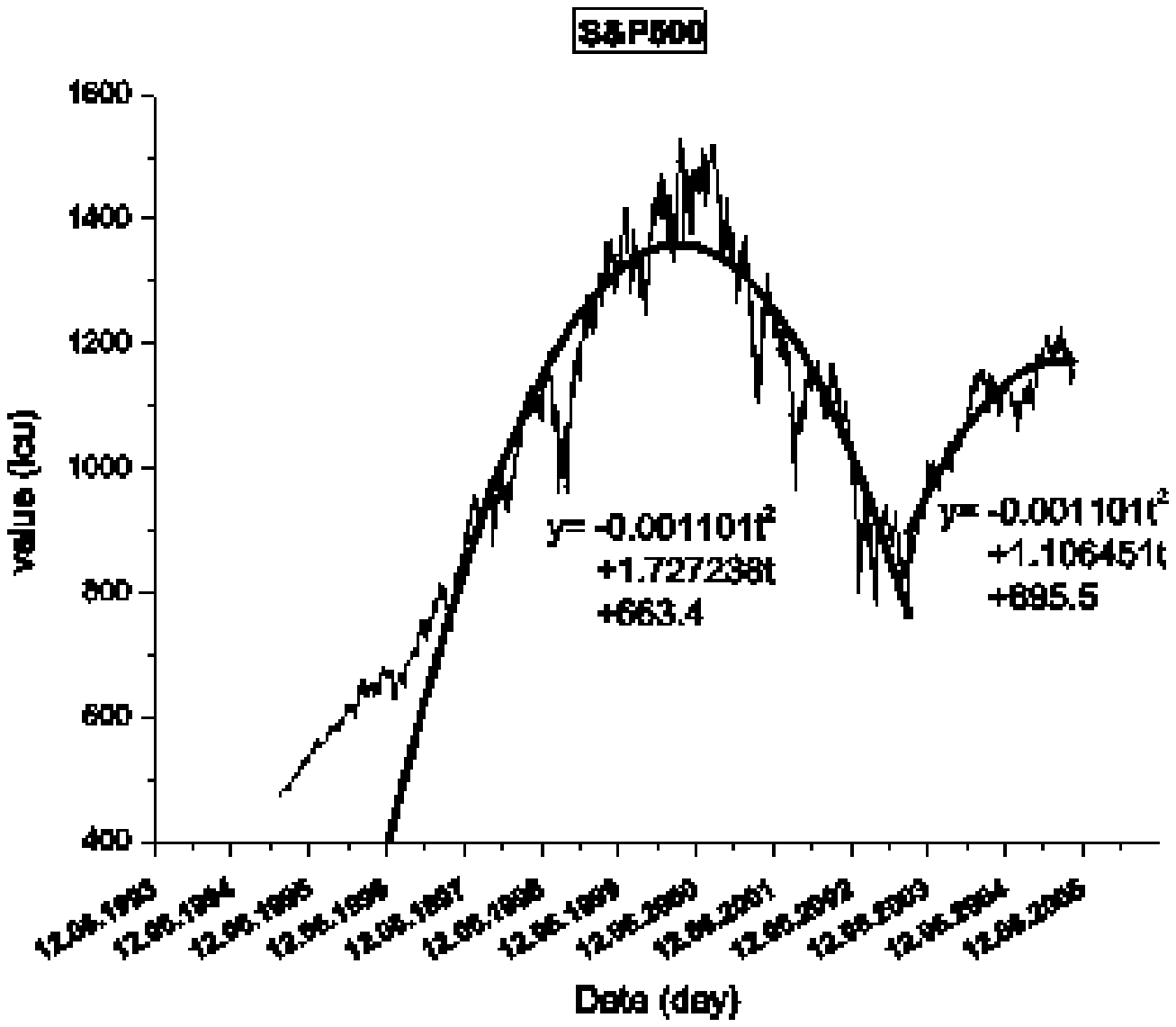}}
\label{fig1}
\end{figure}

Fig. 1.

\begin{figure}[htbp]
\centerline{\includegraphics[width=5.87in,height=5.89in]{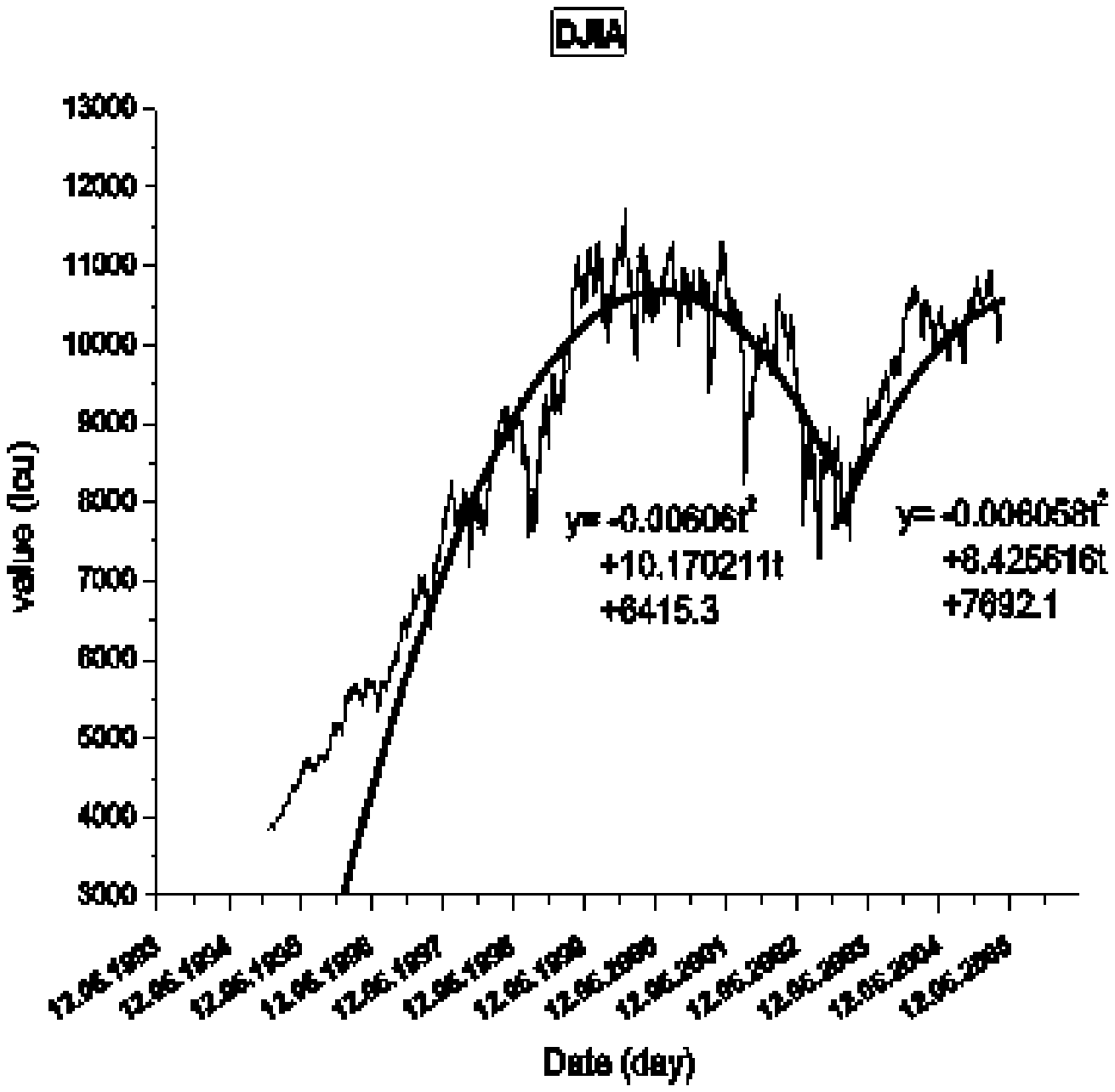}}
\label{fig2}
\end{figure}

Fig. 2.

\begin{figure}[htbp]
\centerline{\includegraphics[width=5.90in,height=5.87in]{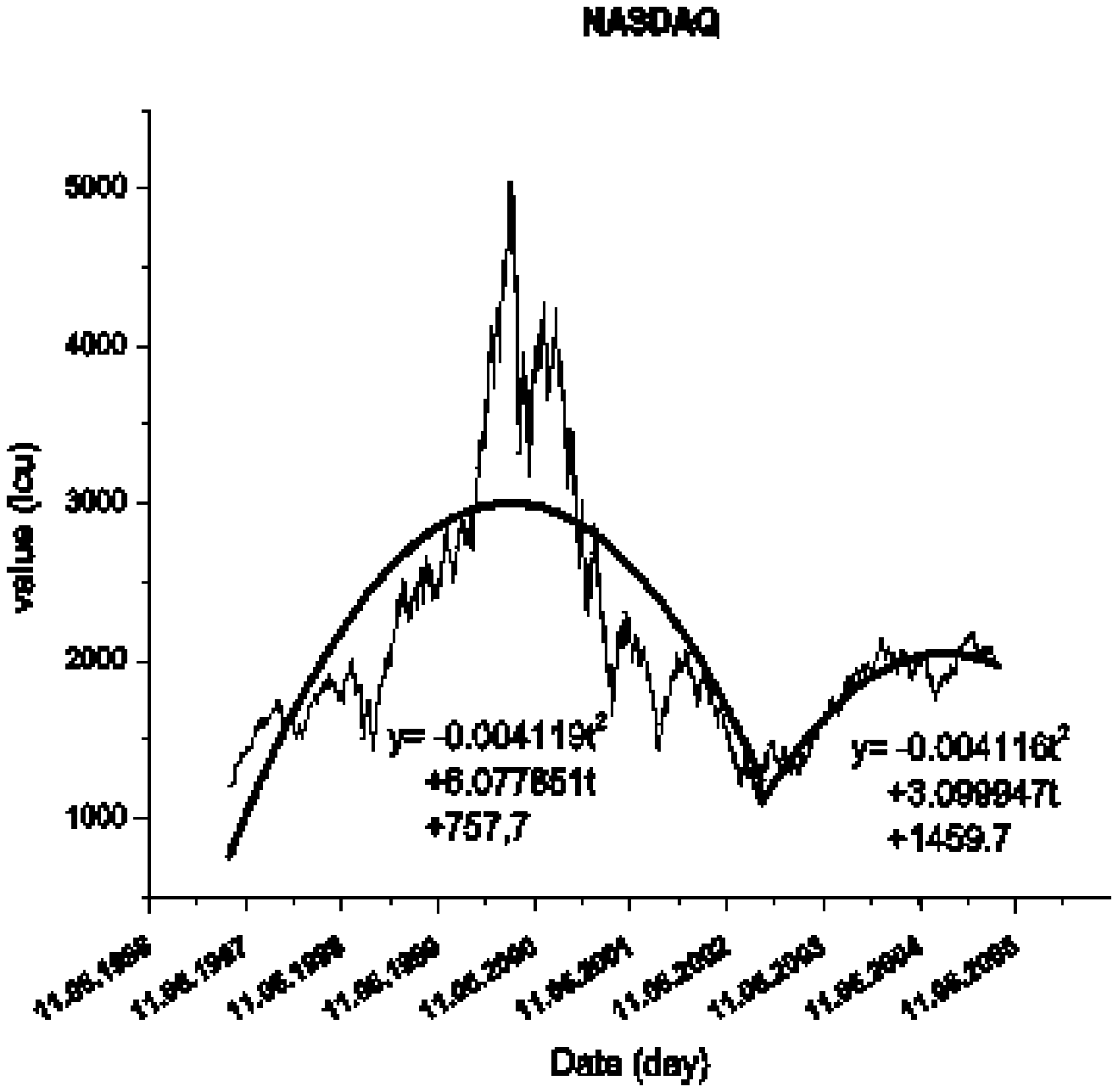}}
\label{fig3}
\end{figure}

Fig. 3.

\begin{figure}[htbp]
\centerline{\includegraphics[width=5.89in,height=5.89in]{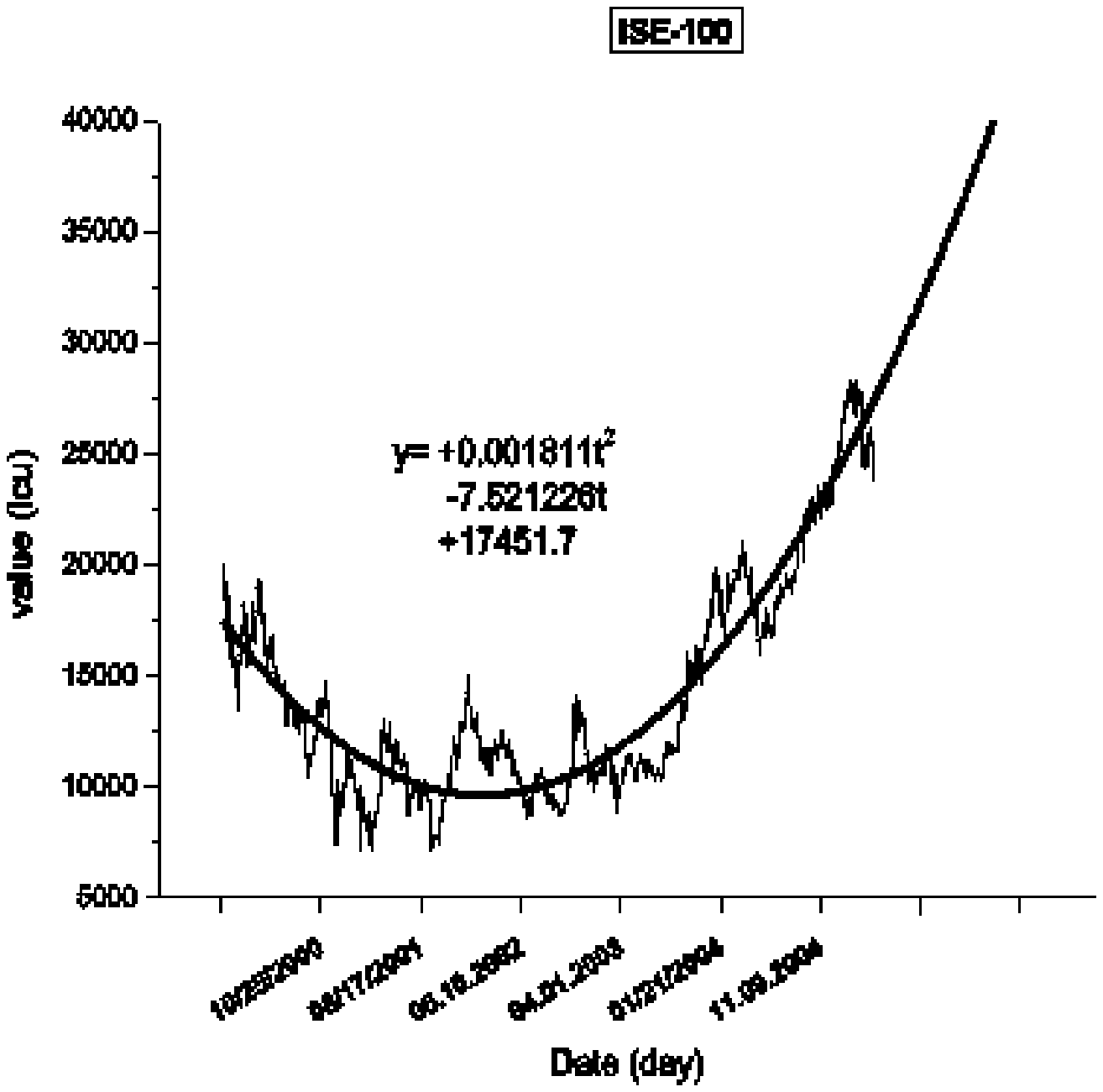}}
\label{fig4}
\end{figure}

Fig. 4. a.

\begin{figure}[htbp]
\centerline{\includegraphics[width=5.89in,height=5.89in]{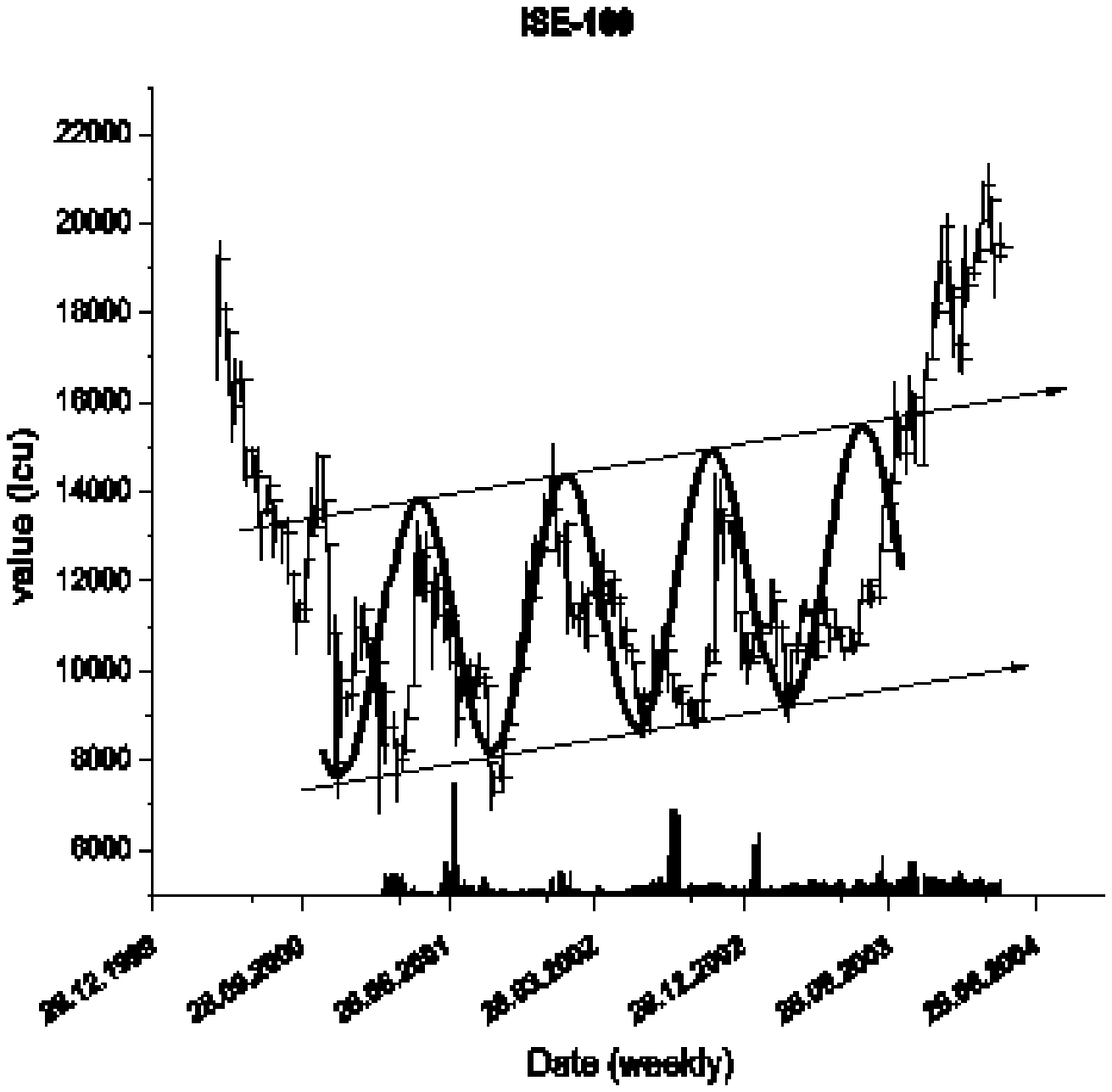}}
\label{fig5}
\end{figure}

Fig. 4. b.

\begin{figure}[htbp]
\centerline{\includegraphics[width=5.89in,height=5.27in]{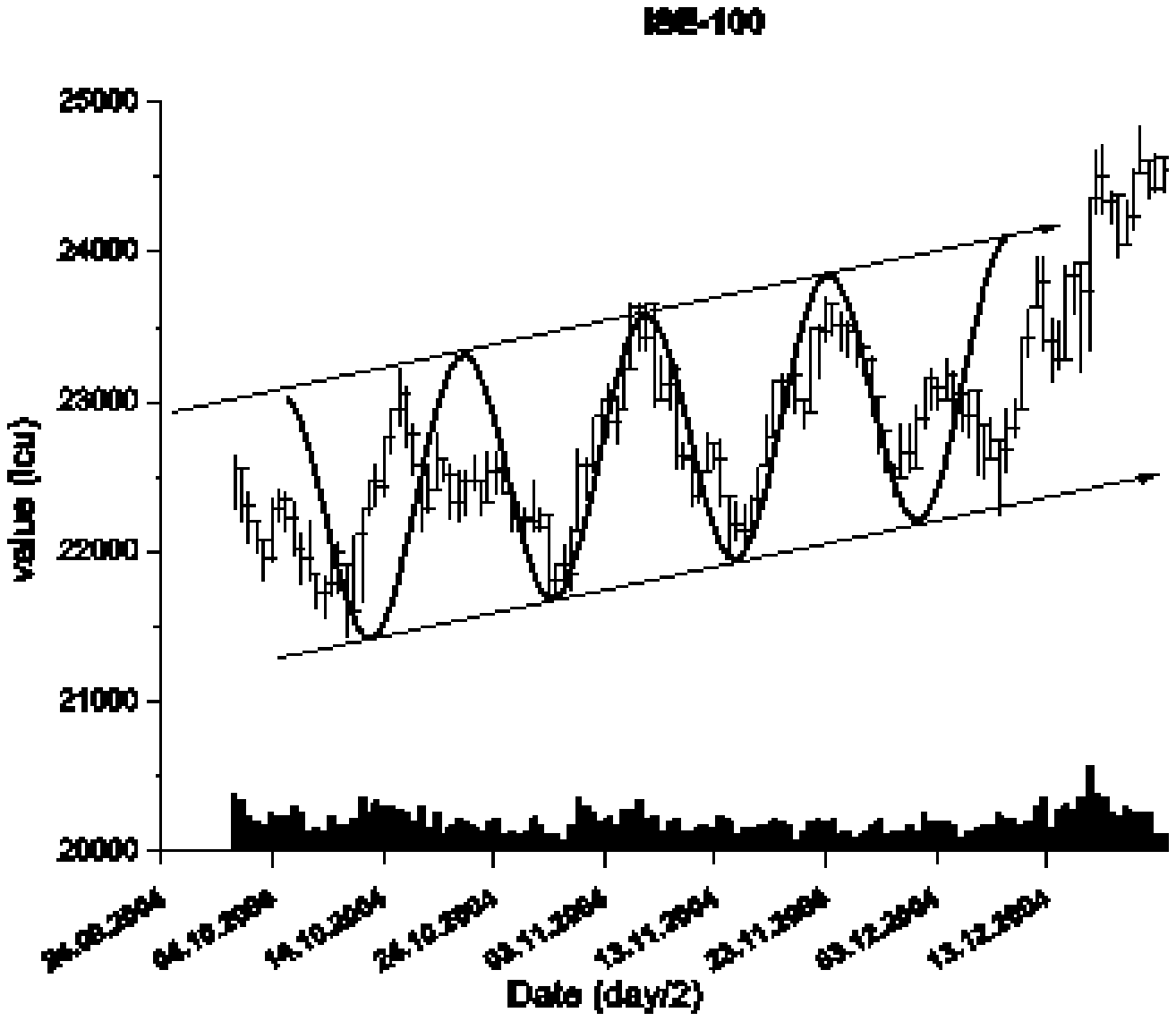}}
\label{fig6}
\end{figure}

Fig. 4. c.

\end{document}